
\newif\iffigs\figsfalse
 \figstrue

\newif\ifamsf\amsffalse

\input harvmac
\iffigs
  \input epsf
\else
  \message{No figures will be included. See TeX file for more
information.}
\fi
\ifx\answ\bigans\else
  \message{This does not look very nice in little mode.}
\fi

\Title{\vbox{\hbox{CLNS 96/1447}\hbox{CU-TP-803}
\hbox{\tt hep-th/9612181}\vskip -.5in }}
{Small Volumes in Compactified String Theory}
\centerline{Brian R. Greene$^\star$ $^\ddagger$ and Yakov Kanter$^\dagger$}
\vskip.3in
\centerline{\hbox{
\vtop{\hsize=1.25in\it
\centerline{$^\star$Department of Mathematics and Physics}
\centerline{Columbia University}
\centerline{New York, NY 10027}
}\hskip 1.4in
\vtop{\hsize=1in\it
\centerline{$^\dagger$F.R. Newman Laboratory}
\centerline{of Nuclear Studies}
\centerline{Cornell University}
\centerline{Ithaca, NY 14853}
}
}}

\footnote{}{$^\star$Email: {\tt greene@math.columbia.edu}}
\footnote{}{$^\dagger$Email: {\tt yakov@hepth.cornell.edu}}
\footnote{}{$^\ddagger$ On leave from F. R. Newman Laboratory of
Nuclear Studies, Cornell University, Ithaca, NY, 14853}
\vskip 1.2in

We discuss some of the classical and quantum geometry associated
to the degeneration of cycles within a Calabi-Yau compactification.
In particular, we focus on the definition and properties of
quantum volume, especially as it applies to identifying the physics
associated to loci in moduli space where nonperturbative effects
become manifest. We discuss some unusual features of quantum
volume relative to its classical counterpart.

\vskip .3in

\noblackbox

 \font\eighti=cmmi8
\font\eightsy=cmsy8 
  \skewchar\eighti='177

\skewchar\eightsy='60

\font\bigrm=cmr10 scaled \magstephalf
\def\inbar{\,\vrule height1.5ex width.4pt depth0pt}
\font\cmss=cmss10 \font\cmsss=cmss8 at 8pt
\def\BZ{\relax\ifmmode\mathchoice
{\hbox{\cmss Z\kern-.4em Z}}{\hbox{\cmss Z\kern-.4em Z}}
{\lower.9pt\hbox{\cmsss Z\kern-.36em Z}}
{\lower1.2pt\hbox{\cmsss Z\kern-.36em Z}}\else{\cmss Z\kern-.4em Z}\fi}
\def\IC{\relax\hbox{$\inbar\kern-.3em{\rm C}$}}
\def\IP{\relax{\rm I\kern-.18em P}}
\def\IQ{\relax\hbox{$\inbar\kern-.3em{\rm Q}$}}
\def\IR{\relax{\rm I\kern-.18em R}}
\ifamsf
  \font\bbl=msbm10
  \def\BZ{\hbox{\bbl Z}}
  \def\IC{\hbox{\bbl C}}
  \def\IP{\hbox{\bbl P}}
  \def\IQ{\hbox{\bbl Q}}
  \def\IR{\hbox{\bbl R}}
\fi

\def\Tr#1{\hbox{{\bigrm Tr}\kern-1.05em \lower2.1ex \hbox{$\scriptstyle#1$}}\,}

\def\tilde{\widetilde}

\Date{12/96}

\newsec{Introduction}

Our  understanding of quantum geometry ---
the geometrical structure underlying quantum string
theory --- has deepened significantly over the past
few years. From our realization of new geometrical
properties associated with an extended object
even at the classical level \ref\rWittenphases{E. Witten,  
{\it Phases of 
$N=2$ Theories In Two Dimensions}, Nucl. Phys. {\bf B403} (1993) 159}
\ref\rTopchange{P. Aspinwall,  B. Greene and
D. Morrison,  {\it Calabi-Yau Moduli Space, Mirror Manifolds and
Spacetime  Topology Change in String Theory}, Nucl. Phys.{\bf B416}
(1994)  414  }, to the numerous
new properties \ref\r{J. Polchinski, {\it TASI lectures on D-branes}, 
hep-th/9611050, to appear in {\it TASI-96 proceedings} and
references therein; P. Aspinwall, {\it K3 Surfaces and String Duality},
hep-th/9611137, to appear in {\it TASI-96 proceedings} and
references therein; S. Shenker {\it Another Length Scale in String
Theory?}, hep-th/9509132 and references therein;
B. Greene, {\it Lectures on Quantum Geometry}, 
to appear in {\it TASI-96 proceedings} and
references therein; J. Schwarz, {\it
Lectures on Superstring and M Theory Dualities},  hep-th/9607201, to
appear in {\it TASI-96 proceedings} and references therein.} 
which the recent progress in nonperturbative
string theory has allowed us to discern, quantum geometry
has shown itself to be a remarkably rich structure. Even
with the impressive progress that has been made, though, there
are still some rather basic aspects which have not been fully
understood. In this letter we focus on one such property which surrounds
an aspect of {\it quantum volume}. This is a topic discussed
at some length in \ref\rsmalldist{P.  Aspinwall, B. Greene, and
 D. Morrison, {\it Measuring Small Distances in N=2 Sigma 
Models}, Nucl. Phys. {\bf B416} (1994) 414 }\  with an interesting follow
up being \ref\rAspinwall{P. Aspinwall, {\it Minimum Distances in
 Non-Trivial String Target Spaces}, Nucl. Phys. {\bf B431} (1994) 78-96 }.

A central theme running through many of the most important
recent developments is the study of string compactifications
in the vicinity of degeneration points in moduli space.
At such points, degrees of freedom which normally have inconsequential
effects on low energy dynamics play a dominant physical role.
It is clearly important, then, to understand where such points
actually occur in the quantum mechanically corrected moduli space.
The relationship with quantum volume arises because 
these points are associated with the collapse of nontrivial 
cycles to ``zero quantum volume''; this oft-used phrase 
certainly deserves meaningful definition and study.

Whenever discussing the extension of
a classical concept to the quantum domain there is an inherent
ambiguity as many quantum concepts can have the same classical limit.
Without specifying a physical incarnation, the subtleties involved in defining
quantum volume might amount to nothing more than semantics. 
In \rsmalldist, fundamental string instantons which wrap around holomorphic
 spheres
were used as a probe of two-cycle volumes within  a Calabi-Yau manifold.
This resulted in some interesting observations regarding the identification
of special points in and the overall structure of the quantum Calabi-Yau 
moduli space;
these results  played a role in 
\ref\rWittenMF{E. Witten, {\it Phase Transitions In M-Theory And F-Theory,}
Nucl. Phys. {\bf B471} (1996) 195}\  and \ref\rAspinwallOrb{P. Aspinwall, 
{\it Enhanced Gauge Symmetries and K3 Surfaces,}\ 
Phys. Lett. {\bf B357} (1995) 329 }, for instance.

Two questions were raised but left unanswered in \rsmalldist. The first 
question,
as we discuss below, concerns the properties of the quantum
volume of spheres involved in  flop/conifold
transitions away from boundaries in moduli space. The second question
concerns  the extension
of the analysis in \rsmalldist\ 
to address the quantum volume of cycles whose dimension is
greater than two. In \rsmalldist\ no procedure was proposed beyond the
seemingly reasonable assumption that if lower dimensional
submanifolds within some chosen cycle have nonzero volume, then the chosen
cycle itself must have nonzero volume as well. Without a direct physical
probe of higher dimensional cycles, it was difficult to proceed further.
In this paper, using more recent developments, we are able to address
each of these issues and reach conclusions that are rather different than what
might have  been anticipated.

With regards the first question, we explicitly calculate the quantum volume
 of rational
curves in flop/conifold transitions.  In \rsmalldist\ this was
done while holding all other moduli at infinity, and the result
obtained was zero. It was speculated \rsmalldist\ that zero
volume --- something a bit orthogonal to  the increasingly vague notion
of there being a minimal distance in string theory ---
 was an artifact of pinning the other moduli at infinity in this manner and
that such rational curves would pick up nonzero volume at interior
points on the transition locus. Circumstantial evidence 
in favor of this speculation
was given in \rAspinwall\ by studying the simpler case
of blown-down orbifolds. We find here, though, that the flop case
is different from the orbifold
and maintains zero quantum volume all along the transition
locus in the moduli space. We note that this is also relevant
for the K\"ahler side of conifold transitions
as these occur at the center of flop transitions.

With regards the second question,
the discovery of higher dimensional structures in the form of
$D$-brane degrees of freedom does provide us now with such a physical probe
of higher dimensional cycles.
Part of our purpose is to
study  some aspects of the picture
which emerges if we  take wrapped
D-brane masses  as the operational definition
of quantum volume.
It is worth noting that as is already apparent from string dualities such as
$R \rightarrow 1/R$, the geometry which emerges in studying some
configuration depends at
least in part on {\it which} probe 
one uses. Thus, there is probably no unique notion of
quantum geometry but rather a spectrum of possibilities associated
with the different ways it is accessed.
  We note, for instance, that using wrapped $D$-brane configurations is
 different from
using D-brane scattering as probes of sub-stringy geometry
\ref\rdouglasshenker{ M. Douglas, D. Kabat, P. Pouliot and
S. Shenker,\enskip  {\it 
D-branes and Short Distances in String Theory}, hep-th/9608024   },
and understanding the detailed relation
between the two would be valuable.

Using this approach we find a general picture in which contrary to one's
classical intuition the assumption made in \rsmalldist\ is not 
true: in quantum geometry the collapse of
a cycle $B$ to a point, with zero quantum volume, does {\it not} necessarily
imply that subcyles of $B$ with lower dimension  
are necessarily squeezed to zero quantum volume
as well. This clearly has bearing on the identification of resulting zero
mass states at the degeneration point since branes wrapping the lower
dimensional cycles will not become massless. Similarly, strings arising
from wrapping a $p$-brane  around such $p-1$-cycles will not become
tensionless, as well.
Such investigations also allow us to clarify the mirror K\"ahler
interpretation of complex structure degenerations of a Calabi-Yau. The
rough statement that the mirror of collapsing $S^3$'s is collapsing
$S^2$'s
is generally
incomplete since all of the even dimensional cycles on the mirror side can
be involved
\ref\rVafadelPezzo{ M. R. Douglas, S. Katz and  C. Vafa, {\it  Small
Instantons, del Pezzo Surfaces and Type I Theory}, hep-th/9609071 }
\ref\rSeibergdelPezzo{  D. Morrison and N.  Seiberg, {\it  Extremal
Transitions and Five-Dimensional Supersymmetric Field Theories},
hep-th/9609070 }
\ref\CGGK{T.M. Chiang, B. Greene, M. Gross and Y. Kanter, \enskip
 {\it Black Hole
Condensation and the Web of Calabi-Yau Manifolds }, Nucl. Phys. B
(Proc. Suppl.) 46 (1996) 82. }. 
 A by-product of
our  discussion is
a procedure for making such identifications precise.

In section 2 we present an explicit calculation of the volume
of rational curves involved in flop or conifold transitions in
a Calabi-Yau three-fold. This mirror symmetry calculation
measures volumes as probed by fundamental string instantons. In section 3
we discuss various ways of defining volume more generally and
describe the approach of using wrapped $D$-brane configurations.
For two-branes, this makes direct contact with the calculation of section 2.
In section 4 we outline the general picture which emerges from
this approach. We first apply our general picture to the case
of flop/conifold transitions, affirming our explicit calculations of
section 2. We then discuss how
on general grounds,  the collapse of
a higher dimensional cycle does not entail the collapse of
lower dimensional subcyles, illustrating a novel feature
of quantum geometry.

\newsec{Quantum Volumes of Flops and Conifolds}
Classical geometry tells us that topology
changing flop  and conifold transitions involve $S^2$'s which shrink
to zero volume and are then replaced either with other $S^2$'s or
with $S^3$'s. Quantum mechanically one can imagine that this classical
statement is modified in some manner. For instance, transitions
might occur at non-zero classical volume which turns out to have
zero quantum volume; or non-zero quantum volumes might play a role.
For the case of flop transitions, this was first studied in
\rsmalldist\ along a calculationally amenable locus in
Calabi-Yau moduli space: a one-parameter subspace along which
only the volume of the shrinking $S^2$  changes  while all other
Calabi-Yau K\"ahler moduli are held fixed at infinity. By using mirror
symmetry, the Picard-Fuchs equation along the mirror of
this  locus, which governs the behavior of periods of
the holomorphic three-form $\Omega$ was found to be
\eqn\eFlop{\left\{(z {d \over dz})^2 - z (z {d \over dz})^2 \right\}(
\int_{\gamma} \Omega ) = 0. }
The regular and logarithmic monodromy solutions yield 
the complex volume for the $S^2$
\eqn\eBIJ{ B + i J = {1 \over 2 \pi i} \log(z). }
In this expression,  $z$ is a local
moduli space coordinate normalized so that $z = 1$ is the flop point.
We see, then, that at $z = 1$ the quantum volume, like the classical
volume, vanishes identically.

Although attaining zero quantum volume, the expectation of
\rsmalldist\ was that this was an artifact of holding all
other K\"ahler moduli at infinity. As mentioned above,
at least in the case of orbifold singularities, there
are examples which bear this out  \rAspinwall.  Presently, we explicitly 
study the flop/conifold case. One expects this case
to be significantly more
difficult than those studied in \rsmalldist\ and \rAspinwall\
since the locus of interest is generally along an interior transition wall
in the moduli space, rather than along a boundary divisor.

To set up the calculation,
let $M$ be a Calabi-Yau threefold with mirror $W$. Following the
by now standard discussion of 
 \ref\rCan{P. Candelas, X. de la Ossa, P. Green and L. Parkes, 
{\it A Pair of Calabi-Yau Manifolds as an
Exactly Soluble Superconformal Theory}, Nucl. Phys. {\bf B359} (1991)
21}
\ref\rMorisonPicardFuchs{D. Morrison, {\it Pichard - Fuchs Equations
and Mirror Maps For Hypersurfaces,} in {\it Essays on Mirror Manifolds I}
(S.T. Yau, editor), International Press, 1992, p.241 }, we can explicitly
evaluate the quantum corrected value of the complexified K\"ahler
class of $M$ by doing {\it classical} geometrical calculations on
$W$ and invoking the mirror map. Abstractly, the nonlinear sigma
model on $M$ --- for fixed complex structure --- depends on
the complexified K\"ahler class $K$. Although the latter is
not a direct physical observable of a given theory, knowledge of
a sufficiently robust set of correlation functions is enough data
to determine $K$ on $M$.
 The volume of a two-cycle $C$ on
$M$ (which we will always take to be an $S^2$) is then given by
$\int_C K$ where $K$ is the complexified K\"ahler form. The latter
is given by
\eqn\eKahler{ K = t_i e^i}
with 
\eqn\eMirromap{t_i = {{\int_{\gamma_i}\Omega \over \int_{\gamma_0}\Omega}},}
$\Omega$ being the holomorphic three-form on $W$, the $\gamma_i$,
$i=1,...,h^{21}_W$ being a basis of three-cycles with log-monodromy
periods at large complex structure, and $\gamma_0$ being a three-cycle
with regular period. As mirror symmetry aligns $H_2(M,\BZ)$ with
the log-period monodromy subspace of  $H_3(W,\BZ)$ (with a similar
statement for their respective cohomological duals), by suitable
change of basis, mirror symmetry identifies the integral
generators  $e_i$ of $H_2(M,\BZ)$ with the $\gamma_i$.
Thus, since $C$ can be written as $C = a^je_j$ and with $e_j$ being a
dual cycle to the class $e^j$, we have
$\int_C K = t_ia^i$. Without loss of generality, then, we
take our curves $C$ to be amongst the  $e_j$ which then have
volume given by $t_j$ as given in \eMirromap. 

To compute quantum volumes of two-dimensional cycles on $M$, therefore,
we need to know
the log-monodromy periods on its mirror $W$. 
 In this section we will explicitly
evaluate these periods for two two-parameter examples which
have Calabi-Yau phases related by a flop transition.
In the first example we exploit an observation made
in \ref\rGMV{ B. Greene, D. Morrison C. Vafa, {\it A Geometric
Realization of Confinement, }  hep-th/9608039 }  that flop/conifold
transitions  of a particular special
sort are subject to strong renormalization resulting in
a significant shift in the location of the flopping wall.
In particular, the wall is pushed out to a toric boundary divisor,
thus fortuitously bringing the explicit calculation under full
analytic control. In the second example, we consider a more standard
type of flop transition involving an interior transition wall. We
analyze two-cycle volumes using a perturbation scheme.
 The discussion
is naturally phrased in terms of toric geometry (see,
e.g. \ref\rFulton{W. Fulton, {\it Introduction to Toric Varieties},
Annals of Math. Studies, vol. 131, Princeton University Press,
Princeton, 1993 } and \rTopchange\  for an introduction  to
this  subject).  

\subsec{Small Resolution of a Singular Quintic in a $\IP^4$}

Consider a family of quintics in a $\IP^4$. This family has one
K\"ahler parameter, i.e. the overall size of the manifold and
$101$ complex structure parameters.
If we deform the complex structure parameters so
that the quintic develops conical singularities as
in \ref\rconifold{P. Candelas, P.S. Green, T. H\"ubsch, {\it Finite
Distance Between Distinct Calabi-Yau Manifolds}, Phys. Rev. Lett. {\bf
62} (1989) 1956; \enskip {\it Rolling Among Calabi-Yau Vacua},
Nucl. Phys. {\bf B330} (1990) 49}, 
 \ref\rGMS{B. Greene, D. Morrison, A. Strominger, {\it
Black Hole Condensation and the Unification of String Vacua },
Nucl. Phys. {\bf B451} (1995) 109  }\ and then perform a small
resolution by blowing up along the $\IP^2$ containing the singular
points, we will get a new family of Calabi-Yaus with 2 K\"ahler
parameters and $86$ complex structure parameters.
The first K\"ahler parameter is the original overall size of the manifold and
the second controls the size of the sixteen homologous
$S^2$'s introduced as the
result of the blow-up. This is the two-parameter model we want to
study. As discussed in \rGMV, this two-parameter model has
two phases related by flops --- with some rather unusual features
which will aid our calculation below.

The mirror of this family has the following toric description. The fan
for the original $\IP^4$ is spanned by $u_1, \ldots , u_5 \in \IC^4$ , where 
\eqn\eUsA{ u_1 = \pmatrix{1 \cr 0 \cr 0 \cr 0},\qquad 
          u_2 = \pmatrix{0 \cr 1 \cr 0 \cr 0}, \qquad 
          u_3 = \pmatrix{0 \cr 0 \cr 1 \cr 0},\qquad
          u_4 = \pmatrix{0 \cr 0 \cr 0 \cr 1},\qquad
          u_5 = \pmatrix{-1 \cr -1 \cr -1 \cr -1} }
Blowing up adds $u_6 = u_4 + u_5$ to the polytope. Thus a typical
manifold $W$ in this two-parameter family is a toric variety
$V_{\Delta}$, where $\Delta$ is the polytope with
vertices $u_1, \ldots , u_6$. 

As we will see shortly, to understand the moduli space of complex
structures of $W$ one needs to consider the total space of the
canonical line bundle on $W$. As a toric variety, this space is built
using polytope $\Delta^{+}$  in $\IC^5$
with vertices given by the origin $O$ of $\IC^5$ and 
\eqn\eVsA{ v_0 = \pmatrix{1 \cr 0 \cr 0 \cr 0 \cr 0}, \enskip
          v_1 = \pmatrix{1 \cr 1 \cr 0 \cr 0 \cr 0},  
          v_2 = \pmatrix{1 \cr 0 \cr 1 \cr 0 \cr 0},
          v_3 = \pmatrix{1 \cr 0 \cr 0 \cr 1 \cr 0},
          v_4 = \pmatrix{1 \cr 0 \cr 0 \cr 0 \cr 1},
          v_5 = \pmatrix{1 \cr -1 \cr -1 \cr -1 \cr -1},
          v_6 = \pmatrix{1 \cr -1 \cr -1 \cr -1 \cr 0}  }

To analyze the phase structure of our moduli space we need the kernel
of $\chi$, which is a 5 by 7 matrix with columns $v_0, \ldots , v_6$.
One can check that a possible basis for the kernel is $\pmatrix{\xi \cr \eta}$,
where
\eqn\eXiEtaA{ \pmatrix{\xi \cr \eta} = \pmatrix{ -5 & 1 & 1 & 1 & 1 & 1 & 0 \cr
                                                -1 & 0 & 0 & 0 & 1 & 1 & -1 }}
The secondary fan of $W$  is generated by the two-dimensional
column-vectors of \eXiEtaA, and  is shown in Fig. 1.

\midinsert
$$\vbox{\centerline{\epsfxsize=3.5in\epsfbox{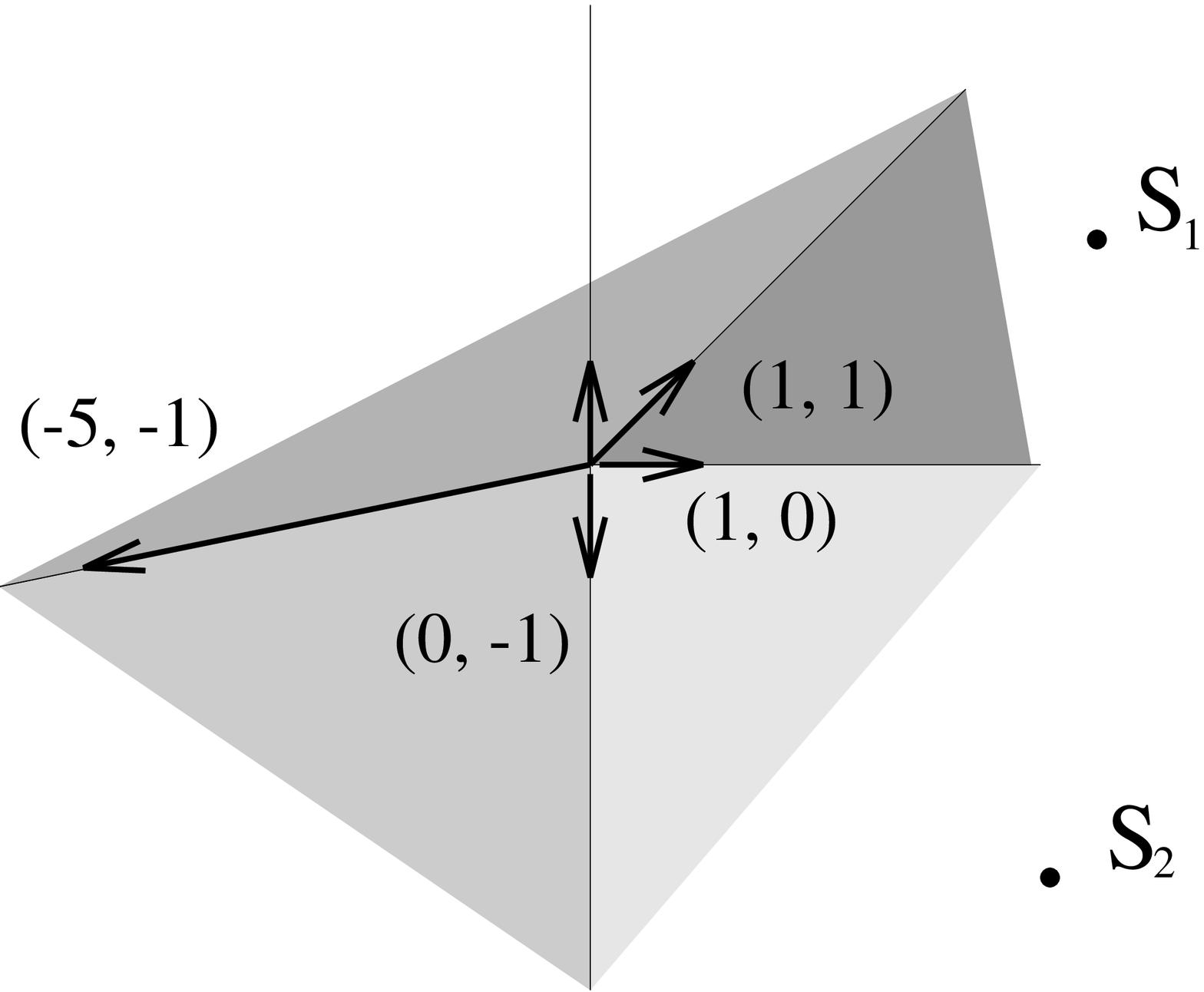}}
\centerline{ Figure 1. The secondary fan of a blown-up quintic in $\IP^{4}$ }
}$$
\endinsert

The deep interior of phase I corresponds to the
neighborhood of the infinite size point of $M$ (symbolically represented
 by the point $S_1$ in Fig. 1) or  the neighborhood of infinite
complex structure of its mirror $W$ . Likewise, the deep interior of phase
II (the neighborhood of $S_2$ in Fig. 1)
 has a similar significance for a different Calabi-Yau $\tilde{M}$,
birational to the first. As explained in \rGMV, $\tilde{M}$ can be obtained
from $M$ by flopping a
complex curve. The mirror of this transition involves  deforming  the
complex structure of  $W$ until a three-cycle vanishes. At this point
in the moduli space of complex structures the space becomes
singular. One can then continue to deform the complex structure
thereby passing once again to a smooth
complex structures on  $W$. So, one
possible approach to the problem of computing the quantum volume of a
two-cycle ${\cal C} \in H_{2}(M,\BZ)$ would be to explicitly
identify its  mirror three-cycle $\gamma \in H_{3}(W,\BZ)$ and
evaluate $\int_{\gamma}\Omega$ with properly normalized three-form
$\Omega$. This is technically difficult. Instead, we can use the
following familiar but less direct method.

$\bullet$ All periods $\int_{\gamma}\Omega$ satisfy Picard-Fuchs
equations. Gel'fand and collaborators \ref\rGKZ{I. Gel'fand,
M. Kapranov, A. Zelevinski, {\it Discriminants, Resultants and
Multidimensional Determinants}, Birkh\"auser, Boston, 1994 } showed
how to write such partial differential equations 
for any toric variety. For our two-parameter example the
Picard-Fuchs system will consist of the following two equations:
\eqn\ePFa{\{ \theta_1^3 (\theta_1 - \theta_2) - x (4 \theta_1 + \theta_2
+ 1) \cdots (4 \theta_1 + \theta_2 + 4)\} \Phi(x,y) = 0}
and
\eqn\ePFb{\{\theta_2^2 + y(4 \theta_1 +\theta_2 +1)(\theta_1 -
\theta_2)\} \Phi(x,y) = 0,} where  \eqn\ethetas{\theta_1 = x \space
\partial_x, \quad \theta_2 = y \space \partial_y} and  
$x$ and $y$ are variables on the space
of complex structures of $W$ such that $x = y = 0$ is the
\hbox{infinite complex structure point}.   \break
$\bullet$ Find  a solution which is single-valued around $x = y =0$
and two solutions with log-monodromy at  $x = y = 0$. From the
previous section we know that the latter  will be of the form
$\int_{\gamma_{1}}\Omega$ and  $\int_{\gamma_{2}}\Omega$, where
$\gamma_1$ and $\gamma_2$ are mirrors of the two-cycles. 
Methods to do this were developed in 
\ref\rBV{V. Batyrev, D. van Straten, {\it Generalized Hypergeometric
Functions and  Rational Curves on Calabi-Yau Complete Intersections in Toric
Varieties}, Commun. Math. Phys. 168 (1995) 493} \rAspinwall.
 The single-valued solution is  
\eqn\eSingleA{\Phi_0(x,y) = \sum_{m \ge 0} \sum_{n \le m} {(4 m +
n)! \over (m!)^3 (n!)^2 (m - n)!}x^m (-y)^n,}
 while the log-monodromy solutions are
\eqn\eLogX{ \Phi_1(x,y) = - \log(x) \Phi_0 + ... }
$$\eqalign{ \Phi_2(x,y) & = - \log(- y) \Phi_0 \cr    
+ \sum_{m \ge n \ge 0} &{(4 m +
n)! \over (m!)^3 (n!)^2 (m - n)!}x^m (-y)^n [2 \Psi(n+1) - \Psi(4 m +
n + 1) - \Psi(m + 1 - n)]\cr
- \sum_{m \ge 0}\sum_{n \ge m+1} &{(n -m -1)! (4 m + n)! \over (m!)^3 (n!)^2} 
(-x)^m y^n. } \eqnn\eLogY\eqno{\eLogY}$$
These power series converge in some neighborhood of $x = y = 0$.
Note that when we hold the overall size of the manifold at infinity,
i.e. when $x = 0,$ \enskip  $\Phi_0(0,y) \equiv 1,$ $\Phi_1(0,y)
\equiv \infty$ as
it should  and
$\Phi_2$ specializes to 
\eqn\ePhiSpec{\Phi_2(0,y) = \log{y-1 \over y}, }
which is precisely the expression found in \rGMV\  for the size of the
flopped curve when the other modulus is held at infinity. As explained
there, the effects of string instantons push the conifold point
to $y = \infty$ and thus at the conifold $\Phi_2$ (and  therefore
the quantum area of the complex curve) is equal to zero. 
Since we now have a full solution, we are in 
position to check the conjecture that this zero volume is an artifact
of the other
K\"ahler modulus being infinite. 

The component of the discriminant
locus which passes through the point $x = 0, y = \infty$ is simply
given locally by $|x| < \varepsilon, y = \infty$ and to study the
behavior of the quantum volume, we simply need to analytically continue
the above expressions for $\Phi_0$ and $\Phi_2$ to the neighborhood of
$x = 0, \enskip 1/y = 0.$ 
To do this, let's write $\Phi_0$ and $\Phi_2$ as Barnes integrals
\eqn\eBarnS{ \Phi_0 = \sum_{m \ge 0}{x^m \over (m!)^3} {1 \over {2 \pi
i}} \int_C {\Gamma(z - m) \Gamma(4 m + z + 1) \over \Gamma(z + 1)^2}
y^z dz} and  
\eqn\eBarnY{ \Phi_2 = \sum_{m \ge 0}{(-x)^m \over (m!)^3} {1 \over {2 \pi
i}} \int_C {\Gamma(z - m) \Gamma(-z) \Gamma(4 m + z + 1) \over \Gamma(z + 1)}
(-y)^z dz, }
where $C$ is the contour in the $z$-plane shown in Fig. 2. 
\midinsert
$$\vbox{\centerline{\epsfxsize=2.5in\epsfbox{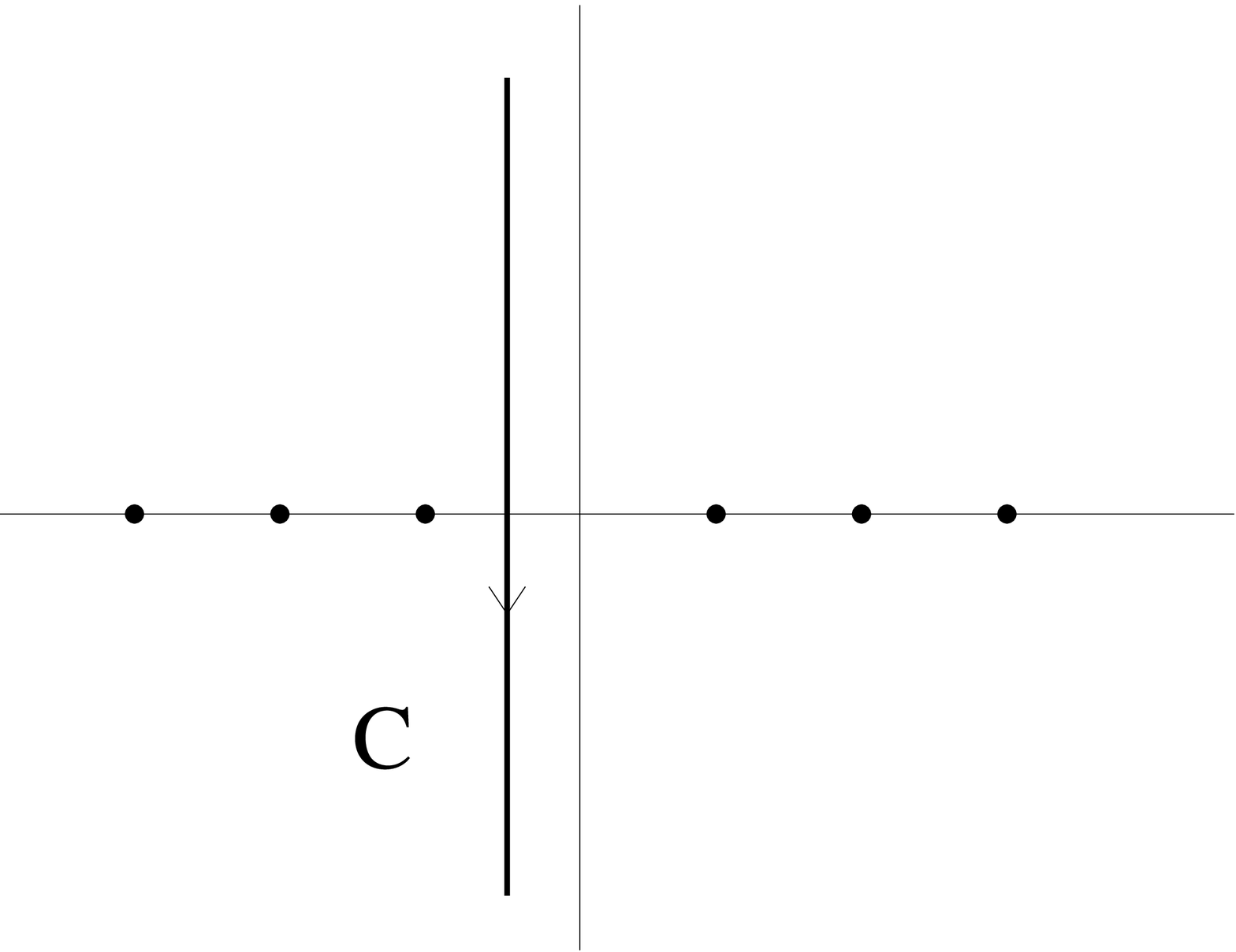}}
\centerline{ Figure 2. Integration contour in equation \eBarnY }
}$$
\endinsert
Closing the contour to the right, we recover our original power series
\eSingleA\ and \eLogY\ which converge for small enough $x$ and $y$. Now
closing the contour to the  left we get power series in $x$ and
$w = {1 \over y}$   which converge for small $x$ and $w$.
We also need, as usual, to specify the branch cuts for our
functions. Recall that the complexified K\"ahler form
 \eqn\eForm{ B + iJ = \sum(B + iJ)_l e^l}
is defined only modulo elements of integral cohomology and therefore
$B_l$ are only defined modulo an integer.
 In our explicit examples we  will cut so that the $0 \le
B_l \le 1$,  which implies $0 \le \arg(x) \le 2\pi$ and   $0 \le \arg(y)
\le 2\pi$.  
 This gives 
\eqn\eLargeY{\Phi_2(x,w) = - \sum_{m \ge 0}(-x)^m \sum_{n \ge 4 m +
1}w^n {\Gamma(n) \over {\Gamma(1 + n + m) \Gamma(n - 4 m)}}. } 
Clearly, $\Phi_2(x,0) \equiv 0$ for small enough $x$ and thus we see
that zero quantum areas persist for manifolds of finite size. 
Note that in this example we are able to compute the volume at the
conifold points exactly due to the fortuitous fact that the branch of the
discriminant locus ${\cal B}$ for which the three-cycles of interest
to  us vanish
is given by a very simple toric boundary equation
\eqn\eDiscrA{(x, y) \in {\cal B} \iff |x| \le \varepsilon, \enskip 1/y
  = 0.}

This feature is  special to  examples in which  the flop
involves the non-compact (line bundle) generator. Therefore it is
important to establish that zero areas also occur in a case of a more
conventional flop.  We now turn to an example of this sort. 

\subsec{Calabi - Yau hypersurfaces in $\IP^4(1, 1, 2, 2, 3)$.}
The general technology here is very similar to the previous example.
As a toric variety, $\IP^4(1, 1, 2, 2, 3)$ is given by the fan which
is spanned by $u_1, \ldots, u_5$ where
\eqn\eUsB{ u_1 = \pmatrix{1 \cr 0 \cr 0 \cr 0},\qquad 
          u_2 = \pmatrix{0 \cr 1 \cr 0 \cr 0}, \qquad 
          u_3 = \pmatrix{0 \cr 0 \cr 1 \cr 0},\qquad
          u_4 = \pmatrix{0 \cr 0 \cr 0 \cr 1},\qquad
          u_5 = \pmatrix{-3 \cr -2 \cr -2 \cr -1}. }  
This manifold is singular and the singularity can be resolved by
adding $u_6 = - u_1 - u_2 - u_3$ to the fan. Now the total space of
the canonical line bundle is given by 
\eqn\eVsB{ v_0 = \pmatrix{1 \cr 0 \cr 0 \cr 0 \cr 0}, \enskip
          v_1 = \pmatrix{1 \cr 1 \cr 0 \cr 0 \cr 0},  
          v_2 = \pmatrix{1 \cr 0 \cr 1 \cr 0 \cr 0},
          v_3 = \pmatrix{1 \cr 0 \cr 0 \cr 1 \cr 0},
          v_4 = \pmatrix{1 \cr 0 \cr 0 \cr 0 \cr 1},
          v_5 = \pmatrix{1 \cr -3 \cr -2 \cr -2 \cr -1},
          v_6 = \pmatrix{1 \cr -1 \cr -1 \cr -1 \cr 0}.  }
The kernel of the matrix $\chi$ with columns spanned by $v_0, \ldots,
v_6$ can be chosen to have the basis $\pmatrix{\xi \cr \eta},$
 where 
\eqn\eXiEtaB{ \pmatrix{\xi \cr \eta} = \pmatrix{ -1 & 1 & 0 & 0 & 1 &
1 & -2  \cr
                                                -3 & 0 & 1 & 1 & -1 & -1
& 3 }.}   
The secondary fan is again generated by the two-dimensional
column-vectors of \eXiEtaB\  and therefore looks as shown in Fig. 3.

\midinsert
$$\vbox{\centerline{\epsfxsize=3.5in\epsfbox{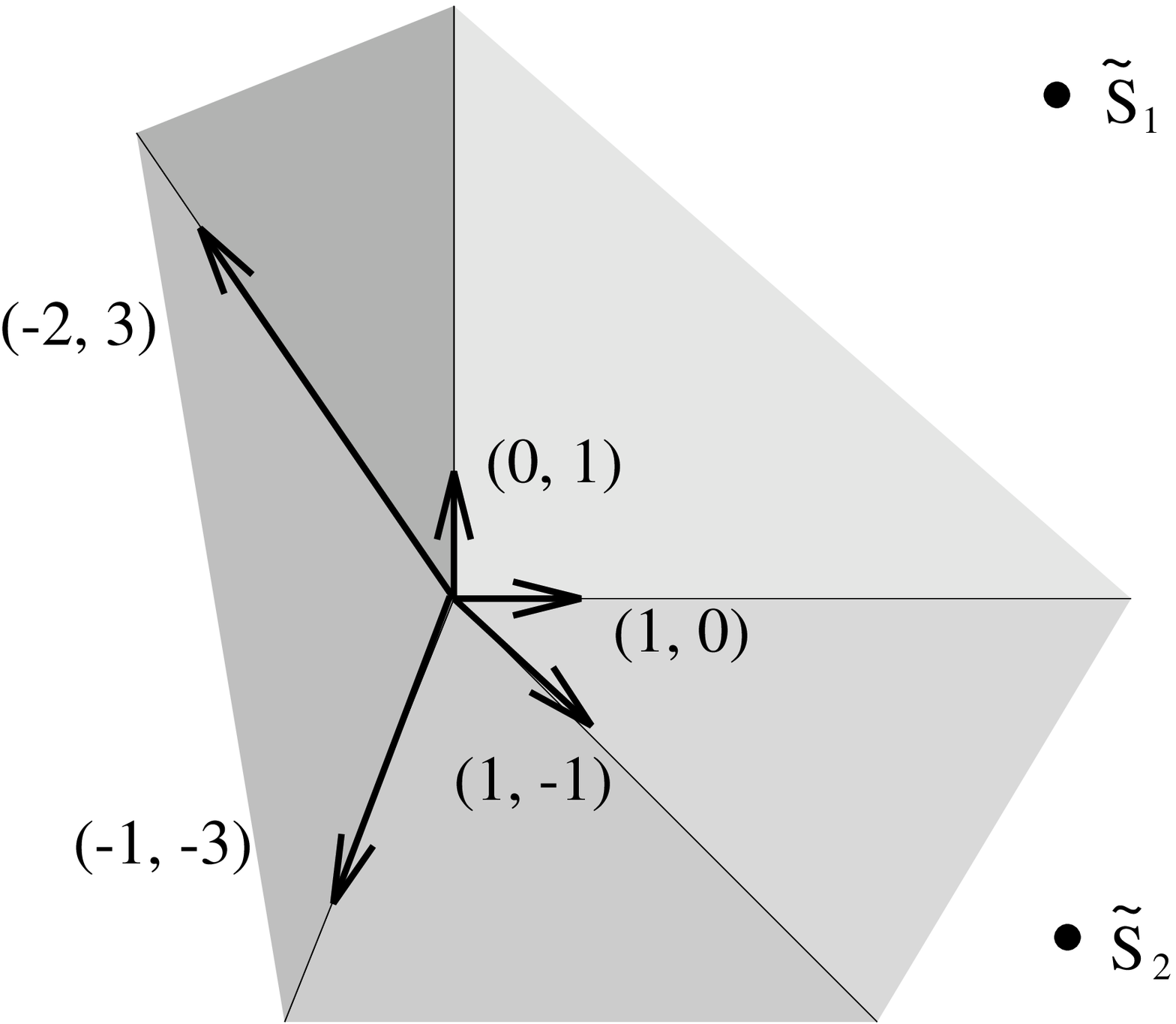}}
\centerline{ Figure 3. The secondary fan of a  Calabi-Yau hypersurface
             in $\IP^4(1, 1, 2, 2, 3)$ }
}$$
\endinsert

We see that again we have two smooth Calabi-Yau phases which differ by
a flop. The Picard-Fuchs system is 
\eqn\ePFaB{\{ \theta_1^2 (3 \theta_1 - 2 \theta_2) ( 3 \theta_1 - 2
\theta_2 - 1) (3 \theta_1 - 2 \theta_2 -2)  - x (\theta_1 -
\theta_2)^2 (3 \theta_1 + \theta_2 + 1) \cdots (3 \theta_1 + \theta_2
+ 3) \} \Phi(x,y) = 0}
and 
\eqn\ePFbB{\{\theta_2 (\theta_2 - \theta_1)^2 - y (3 \theta_1 + \theta_2
 + 1) (3 \theta_1 - 2 \theta_2) (3 \theta_1 - 2 \theta_2 -
1)\}\Phi(x,y) = 0,}
where coordinates $x$ and $y$ are chosen so that $x = y = 0$
corresponds to the infinitely large  complex structure. 
The part  of the discriminant locus we are interested in is given
in parametric form by
$$\eqalign{ y  = & {{(3 - 2 s)^3} \over {(1 - s)^2 (3 + s)^3}} \cr
            x  = & { {s (1 - s)^2} \over {(3 + s) (3 - 2 s)^2} },}
\eqnn\eDL\eqno{\eDL} $$  
where $|s| < \delta$ for some $\delta > 0$.

A single-valued and  log-monodromy solution can be written in terms of
Barnes-type integrals as 
\eqn\eSingB{\Phi_0(x,y) = \sum_{n \ge 0}{(-x)^n \over n! }{1 \over 2 \pi i}
 \int_{\cal C}{\Gamma(z - n) \Gamma(3 z + n + 1) (-y)^z dz \over 
\Gamma(n - z + 1) \Gamma(z + 1)^2 \Gamma(3 z - 2 n +1)}}
and 
\eqn\eLogB{\Phi_2(x,y) = \sum_{n \ge 0}{x^n \over n! }{1 \over 2 \pi i}
 \int_{\cal C}{\Gamma(z - n)^2 \Gamma(3 z + n + 1) y^z dz \over 
 \Gamma(z + 1)^2 \Gamma(3 z - 2 n +1)},}
where contour ${\cal C}$ is shown in Fig. 4. 
\midinsert
$$\vbox{\centerline{\epsfxsize=2.5in\epsfbox{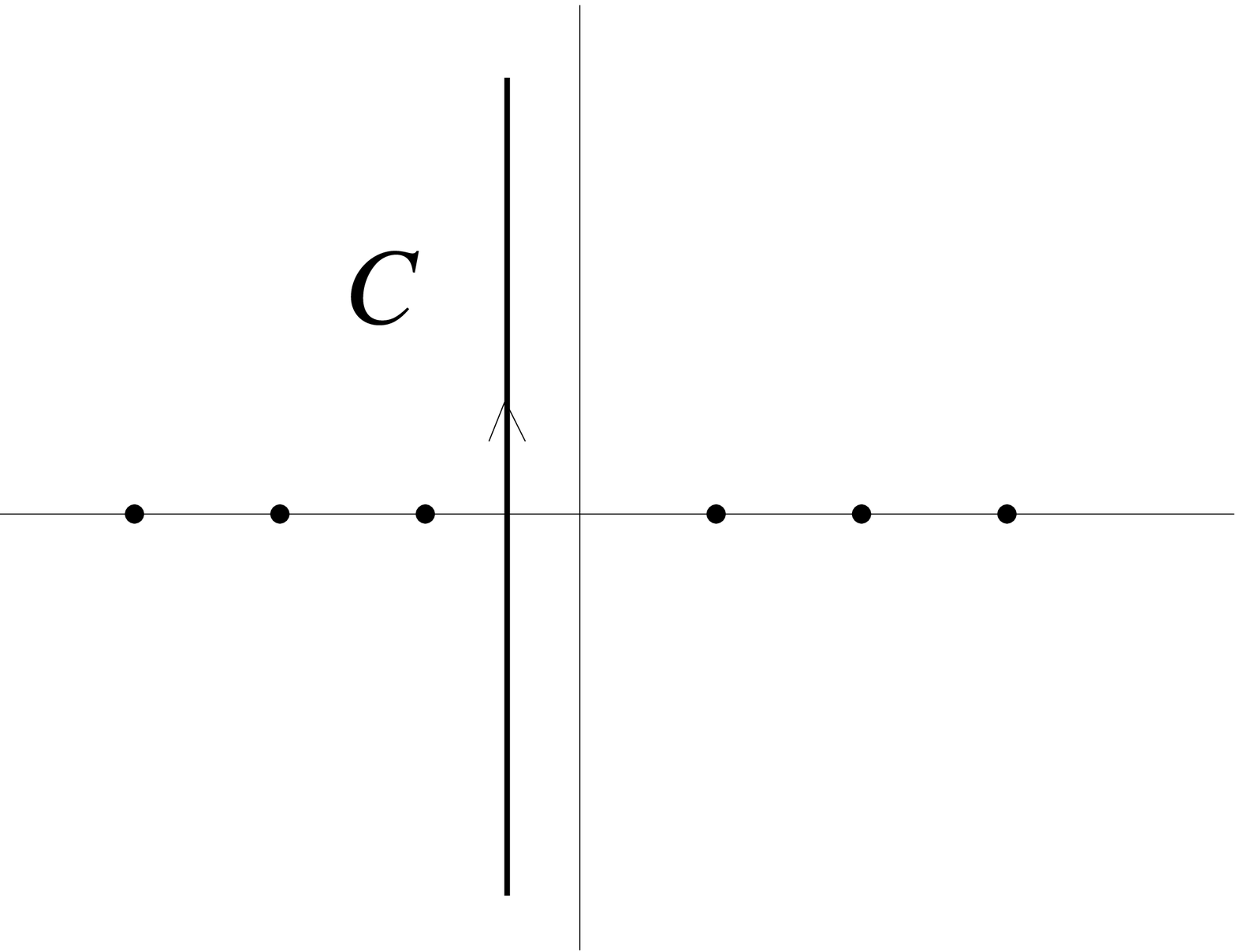}}
\centerline{ Figure 4. Integration contour in equation \eLogB } 
}$$
\endinsert

When the overall size of the manifold is held at infinity (i.e. for
$x=0$), \eLogB\  specializes to $\Phi_2(0,y) = \log(y)$ in 
agreement with results of \rsmalldist\ for a flop. In particular, we
recover the zero area for the flopped 2-cycle for $x=0$ and $y =
1$. However, now we can investigate whether this zero area persists
when we move away from infinite radius point, i.e. for $x \not=
0$. Substituting \eDL\ for the component of  discriminant locus
passing through $x = 0, y = 1$ into \eLogB\ and \eSingB, we find
explicit expressions for the quantum area of our two-cycle on the
discriminant locus. We could not find an easy analytic proof that this
expression is identically zero. However, one can expand $\Phi_2(x(s),
y(s))$ as power series in $s$ around $s=0$, where $s$ is the parameter
of equation \eDL. We
carried out this expansion to the fifth order in $s$ and found that
the coefficients of the power series are zero. This proves (at least
to that order) that the quantum volume of two-cycles remains zero for
manifolds of finite overall size. As we will discuss further in
section four, the fact that $\Phi_2$ is zero on this component of the
discriminant locus should follow on more general grounds
 from the vanishing cycle being an {\it
integral} class, i.e. an element of $H_3(W, \BZ)$.

\newsec{Quantum Volume}

In the above discussion, we have made use of a notion of quantum
volume that has been developed over the last couple
of years in the context of conformal field theory
and mirror symmetry. As such,
it naturally gives us a notion of quantum volume for two-cycles as these
are the geometrical structures which control instanton corrections.
In this section, we briefly examine this notion and its
extension to a definition of quantum volume for higher even-dimensional
 cycles.

There are many ways one can attempt to define a notion of  volume in
string theory. Here
we shall discuss three that are interrelated and in one way or another
have played a prominent role in recent physical developments.

\subsec{Volume from the Linear Sigma Model}

The first comes from the linear sigma model of Witten
\rWittenphases. The linear sigma model
is a physical realization of the symplectic quotient from classical
geometry. In this way, it provides a direct link between classical
and quantum geometry. To keep the discussion simple, let's consider
the example of a degree eight Calabi-Yau
in $W\IP^4(1,1,2,2,2)$. This Calabi-Yau has $h^{11}=2$ and we therefore
represent the linear sigma model on this space by a $U(1)^2$ gauge
theory as follows:

\eqn\eLSM{ {\cal L} = {\cal L}_{\rm kin} + {\cal L}_W + {\cal L}_{\rm
 gauge} + {\cal L}_{\theta} + \sum_{a = 1,2}r_a \int d^2y D_a  }

We will discuss those terms in \eLSM\ that are directly relevant
for our purposes here. A more detailed discussion of this Lagrangian
can be found in \rWittenphases.

Vacuum solutions are determined from the condition that the bosonic
potential energy $U$ is zero. In our case 
\eqn\ePot{ U = {1 \over 2{e_1}^2}{D_1}^2 +{1 \over 2{e_2}^2}{D_2}^2,}
modulo terms whose exact
form is unimportant for our discussion, where
\eqn\eFI{ D_a = -{e_a}^2(\sum_{i = 0}^{6}{Q_i}^a|\phi_i|^2 - r_a),  \qquad
a = 1,2.  } 
As discussed in
\ref\rAG{P. Aspinwall, B. Greene, {\it On the Geometric Interpretation
of N=2 Superconformal Theories}, Nucl. Phys.{\bf B437} (1995) 205-230}, the
charges ${Q_i}^a$ can be determined from the gauge invariance of the
superpotential. A convenient choice is
\eqn\eQs{ \pmatrix{ {Q_i}^1 \cr {Q_i}^2} = \pmatrix{ -8 & 1 & 1 & 2 & 2 &
2 & 0  \cr
                                                0 & 1 & 1 & 0 & 0 & 0
& -2 }.}  
The two Fayet-Illiopoulos D-term parameters $r_1$ and $r_2$ have
a direct classical geometrical interpretation. Abstractly, they
specify the moment map whereby the K\"ahler structure on the
ambient $\IC^6$ induces a K\"ahler structure on the line bundle
${\cal O}(8)$, as discussed in \rWittenphases.  More concretely, we 
see that when $r_1$ and $r_2$ are both positive, vanishing
of the D-terms requires

\eqn\eDtermA{  |\phi_1|^2 +  |\phi_2|^2 + 2|\phi_3|^2 + 2|\phi_4|^2 +
2|\phi_5|^2 - 8|p|^2 - r_1}
\eqn\eDtermB{  |\phi_1|^2 + |\phi_2|^2 - 2|\phi_6|^2 - r_2.}

In the smooth Calabi-Yau phase, $p$ is forced to zero by transversality,
in the usual way, and we directly see that $r_1$ controls the size of
the ambient projective space. The relevance of $r_2$ becomes clear when
we recall that the original weighted projective space is singular at
$\phi_1 = \phi_2 = 0$. For $r_2 > 0$, the second $D$-term  ensures that
$\phi_1$ and $\phi_2$ can not simultaneously vanish. More precisely, when
$\phi_6 \ne 0$, the second $U(1)$ together with its $D$-term (which
yield a $\IC^*$ action) can be used to set $\phi_6 = 1$ and recover
the original Calabi-Yau, except for the singular locus. When $\phi_6 = 0$,
we recover the singular locus (a quartic in $\phi_{3,4,5}$) except
that the second $D$-term replaces the point $\phi_1 = \phi_2 = 0$
(at fixed $\phi_{3,4,5}$) with a $\IC\IP^1$ whose radius is determined
by $r_2$. Thus, this classical analysis clearly shows that $r_1$
controls the size
of the ambient (singular) weighted projective space  $W\IP^4(1,1,2,2,3)$
while $r_2$ controls the size
of the resolving space. 

In the quantum theory (in the two-dimensional field theory sense),
the linear sigma model parameters
will flow via the renormalization group to parameters $\tilde
r_1$ and $\tilde r_2$
describing the conformally invariant nonlinear sigma model.
In the language of \rsmalldist\ the algebraic
measure will flow to the sigma model measure, a feature
that has also played an important role in the phase structure
of $M$ and $F$-theories on Calabi-Yau manifolds
\rWittenMF. One might
therefore say that the original parameters $r_1$ and $r_2$
directly measure classical sizes while 
$\tilde r_1$ and $\tilde r_2$ measure their 
string theoretic counterparts. We are working at
string tree level, and hence our discussion is classical from the string
perspective. If we are in type IIA string
theory, though, the K\"ahler parameters are not renormalized
by quantum string effects and we therefore can take 
$\tilde r_1$ and $\tilde r_2$ to be the parameters directly
measuring the quantum volumes involved. 

There are a few comments we should make. First, the linear
sigma model is only applicable for a limited class of
Calabi-Yau compactifications --- those realizable in a toric
setup.
Thus, this definition of {\it classical} volume in terms of
the original linear sigma model coordinates is somewhat limited and
not fully intrinsic to the Calabi-Yau --- it is dependent, rather, on whether
and how the Calabi-Yau is
being realized in an embedding space. We will return to this point
in a moment. Second, we have not been careful above to distinguish
between two, four and six cycles on a Calabi-Yau. The parameters
$r_2$ and $\tilde r_2$ can be thought of as directly measuring, from
the linear and nonlinear sigma model perspectives, the 
volume of the two-cycle involved in resolving the singularity of
the Calabi-Yau space. But how do we get the volume of the whole Calabi-Yau
from $r_1, r_2$ or $\tilde r_1,\tilde r_2$? Do we use the cubic form
that arises from classical geometry? If not, what is the appropriate
quantum analog? In fact, these two comments are directly related, as we will
momentarily discuss.

\subsec{Volume from The Non-Linear Sigma Model}

A nonlinear sigma model on a Calabi-Yau $M$, is specified by a choice
of complexified K\"ahler form $K = B + iJ$ and complex structure 
for $M$. If $W$ is again the mirror to $M$, then in the neighborhood
of a large complex structure point of $W$, the mirror map determines
$K$ on $M$ to be of the form $K = t_je^j$
\rCan  with
\eqn\emap{t_j = {1 \over 2 \pi i}{\rm log}(z_j) + 
{\cal O}(z_1,...,z_{h^{21}(W)}).}
In this equation, the local coordinates on the complex structure
moduli space of $W$ are chosen so that the large complex structure
point corresponds to all $z_i = 0$. The first term on the right-hand-side
of \emap\ arises from the monomial-divisor mirror map
of \ref\rAGMmdmm{P. Aspinwall, B. Greene and D. Morrison, {\it The
Monomial-Divisor Mirror Map }, Internat. Math. Res. Notices (1993)
319}  and can be thought of as the leading order
term to the mirror map in the limit that $\alpha' \rightarrow 0$.
The other terms in \emap\ arise from instanton corrections. When
there is a linear sigma model representation of $M$ and $W$,
${\rm log}(z_j)$ corresponds precisely to a Fayet-Illioupolos parameter
$r_j + i \theta_j$. This matches well with our discussion above
in which these complexified Fayet-Illioupolos parameters
are associated with classical geometry. The instanton
corrections renormalize these classical parameters to
their quantum values, $t_j$. In particular, world sheet instantons which wrap
around $S^2$'s directly probe these values as  $t_j =\int_{S^2_{j}}K$
is precisely action of such a configuration.

It is worth emphasizing three points. First, if one supplies
the data of a complexified K\"ahler class (and a complex structure)
to define a nonlinear sigma model, the K\"ahler class directly
measures quantum volumes. To reduce to a classical measure
of volume one would need to judiciously take the limit as
$\alpha' \rightarrow 0$. The
linear sigma model and/or mirror symmetry provide systematic
means for doing so.
Second,
the allowed choices for $K$ ---
the quantum K\"ahler moduli space --- are generally different from the
classical moduli space. Some classical regions are simply
unattainable while other classically forbidden regions are required.
Simple examples of the former are regions in one-parameter Calabi-Yau
examples with $Im(K)$ sufficiently small (e.g. on the quintic 
$Im(K) \ge J_0$ with $J_0 = {4 \over 5}\sin^3(2\pi/5)$  
\rCan ) and examples of the latter are flopped phases.
Third, our discussion is at string tree level, but should be thought
of as being in the context of the type IIA string on $M$ or the
type IIB on $W$. In either case, the moduli being discussed lie
in vector multiplets and hence do not receive string loop corrections.
The word quantum --- initially introduced to describe two dimensional
conformal field theory corrections --- can then be interpreted in
the full sense of quantum string theory.

So far then, we have a classical and quantum notion of two-cycle
volumes; the quantum notion being the quantity of true
physical relevance. How can we extend this to higher dimensional cycles?
For two-cycles, the fact that string instantons probe their size gave
us a direct physical probe. For higher dimensional cycles, one naturally
seeks a higher dimensional probe and D-branes fit the bill.

\subsec{Volume from Wrapped D-branes}

In type IIB string theory on $W$, BPS saturated three-branes states
wrapped around supersymmetric three-cycles
\ref\rBBS{K. Becker, M. Becker, A. Strominger, {\it Fivebranes,
Membranes and Non-Perturbative String Theory,}  Nucl. Phys. {\bf B456}
(1995) 130}  have mass given by
\ref\rFerrara{A. Ceresole, R. D'Auria, S. Ferrara and A. Van Proeyen,
{\it Duality Transformations in Supersymmetric Yang-Mills Theory
Coupled to Supergravity,}  Nucl. Phys. {\bf B444} (1995) 92}
\eqn\eBPSmass{M =  g_5 e^{{\cal K}/2}|m^I F_I - n_I Z^I|}
where $g_5$ is a positive constant, ${\cal K} = -
\log(i F_I \overline{Z}^I - i Z^I \overline{F}_I),$ 
\eqn\eF{ F_I = \int_{A_I} \Omega}
\eqn\eZ{ Z^I = \int_{B^J} \Omega} 
for some symplectic basis $\{A_I,
B^J\}$ of $H_3(W, \BZ)$ and 
\eqn\eN{n_I = {1 \over g_5} \int_{A_I \times S^2} {\bf F}}
\eqn\eM{m^J = {1 \over g_5} \int_{B^J \times S^2} {\bf F,}}
where ${\bf F}$ is the RR-charge carrier five form.
 In the special case of, say, a magnetically neutral 
and singly electrically charged state, this formula shows
that the three-brane mass is proportional to the period of
the cycle it wraps. When a cycle collapses,
therefore, we expect a new massless state to appear
in the theory 
\ref\rStrominger{A. Strominger,
{\it Massless Black Holes and Conifolds in String Theory,
} Nucl. Phys. {\bf B451} (1995) 96  }. In this sense, these
three-branes are a direct probe of the geometrical properties
of three-cycles on $W$. As the BPS mass formula is exact and
since the full quantum geometry in this sector of the theory
is captured by lowest order classical calculations, wrapped
three-branes of this sort probe the full quantum geometrical structure.

Of particular importance is the fact proven in \rBBS\ that for an
arbitrary three cycle $\gamma$
\eqn\vol{{\rm Vol}(\gamma) \ge |\int_{\gamma} \Omega|}
with equality being achieved for supersymmetric three-cycles, the ones
giving rise to BPS saturated states. Thus, the mass of wrapped $D$-3-branes
directly tracks three-cycle volumes on $W$.
Now that we have a physical observable which directly probes three-cycle
volumes on $W$, we can use
mirror symmetry, to transport this quantum geometric
understanding to  $M$. In essence, since observables like particle
masses are preserved by mirror symmetry, we define the quantum volumes
on $M$ in terms of wrapped $D$-brane masses on $M$ --- which we
directly compute from wrapped $D$-brane masses on $W$. Quantum volumes
on $W$ are thereby taken to quantum volumes on $M$.

 To do so we first recall that
the homology $H_3(W,\BZ)$ is mirror to the sum of  even cycles
$H_0(M,\BZ) \oplus H_2(M,\BZ) \oplus H_4(M,\BZ) \oplus H_6(M,\BZ)$
on $M$
\ref\rAL{P.Aspinwall and C. Lutken {\it Geometry of Mirror Manifolds},
Nucl. Phys. {\bf B355} (1991) 482.}
\ref\rM{D. Morrison, {\it Making enumerative predictions by means of
mirror symmetry}, alg-geom/9504013. }. In particular, the three-cycles
$\gamma_i$, $i=1,...,h^{21}_W$
in $H_3(W,\BZ)$ with
logarithmic periods at infinite complex structure --- points of
maximal unipotent monodromy --- are mirror to $H_2(M,\BZ)$.
Thus, using wrapped two-branes on $M$ or wrapped three-branes
(wrapping three-cycles with the stated monodromy) on $W$ we
have states of mass
\eqn\eMasslog{ M = {{ |\int_{\gamma_i} \Omega| \over (\int_W 
\Omega \wedge \bar \Omega)^{1 \over 2} }}. }
Up to a normalization factor
 $N = {{(\int_W \Omega \wedge \bar \Omega)^{1 \over 2}
\over \int_{\gamma_0} \Omega}}$, this is the  absolute value of the
 same formula obtained
earlier for the quantum volume of two-cycles on $M$, probed with
string instantons. Thus,
using $N$ as our conversion factor from string instanton
to two-brane probes, fully corrected
two-branes wrapping two-cycles (as gotten from mirror symmetry)
and string instantons measure the same quantum two-cycle volumes. 
In particular,
so long as $N$ is well behaved, strings and two-branes agree on
when a two-cycle collapses. Thus, for instance, in
the explicit calculations done in section 2, the center of a flop
which has zero quantum volume (using string instanton probes) also
has zero quantum volume using wrapped two-branes, and thus corresponds
to a point with new massless states. This is one kind of mirror
partner to Strominger's discussion of collapsing three-cycles
in \rStrominger.

This discussion naturally leads us to consider the other three-cycles in
$H_3(W,\BZ)$ which do not have logarithmic monodromy periods at infinity.
These three-cycles are mirror to the other even dimensional
cycles on $M$ and following the same prescription as above,
give us a definition for their quantum volume as well. Namely,
consider an even dimensional integral cycle $C_{\rm even}$
on $M$, with mirror the integral three-cycle $\gamma_{C_{\rm even}}$ on
$W$. Then, we take the quantum volume of $C_{\rm even}$ to be
$N | {{\int_{\gamma_{C_{\rm even}}} \Omega \over
 \int_{\gamma_0} \Omega}} | $. At large complex structure,
this reduces to the classical volume as measured with the 
K\"ahler form $K$, but differs from it at other points in the moduli space.
By construction, this definition also has the virtue of giving
zero quantum volume at those points in the moduli space where we
expect new massless states to arise.
As in the discussion of
two-cycles here as well as in \rsmalldist \rAspinwall, this definition
of quantum volume is not monodromy invariant ---
this, for instance, accounts for the cuts required in
section 2. Rather, it is the whole
tower of BPS states which is invariant while any individual state
transforms nontrivially amongst the others. Although different from
what we are familiar with in classical geometry, this  path dependence
in the moduli space is a basic feature of quantum geometry.
 We briefly indicate some of the implications of this in the next section.

\newsec{General Picture}

The discussion of the previous sections leads to the following general
picture. Let $\cal M$ be the moduli space of complex structures
on $W$ (for, say, fixed and large K\"ahler class). We show this
in figure 5.
\midinsert
$$\vbox{\centerline{\epsfxsize=4in\epsfbox{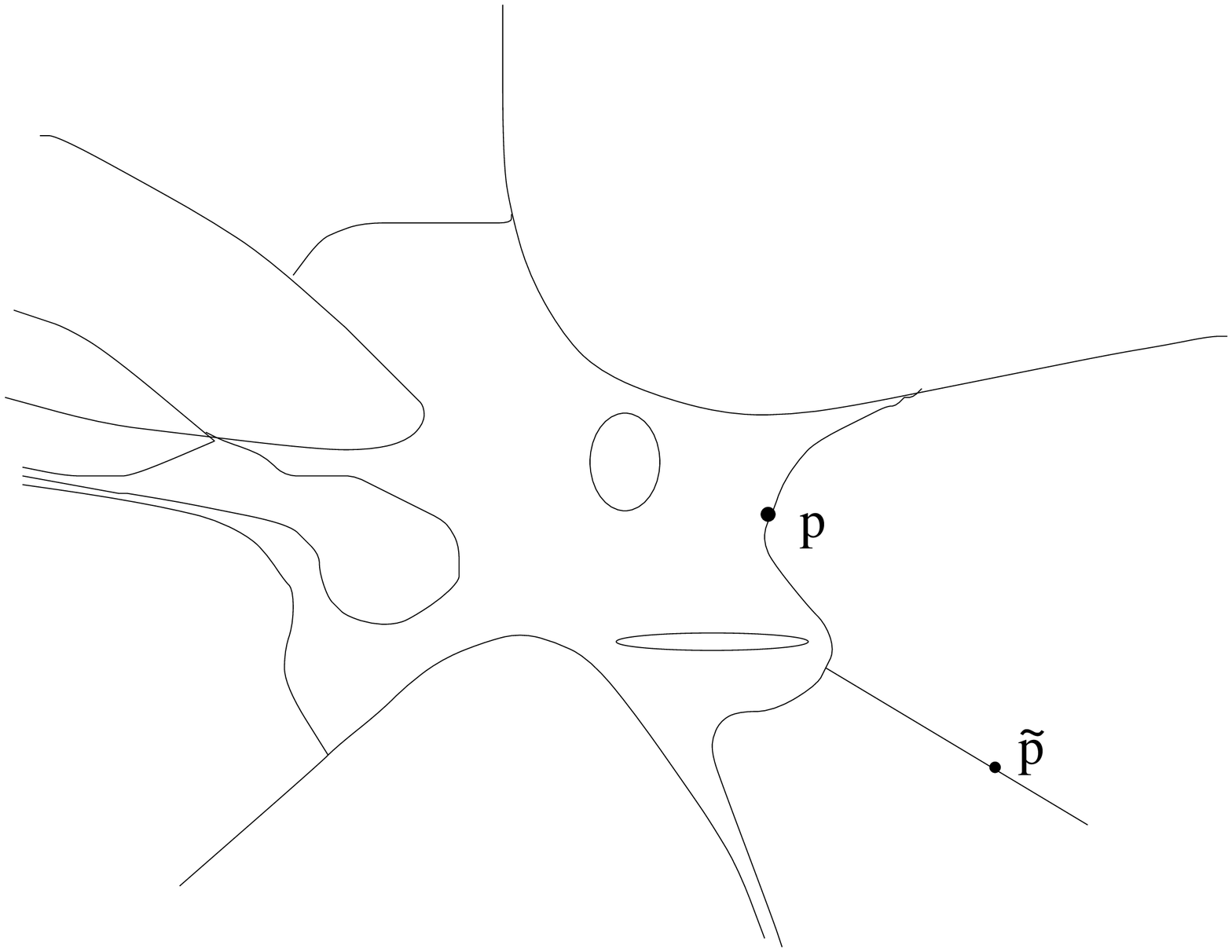}}
\centerline{Figure 5. A typical discriminant locus within a 
 complex structure moduli space }
}$$
\endinsert
 The curves meandering through $\cal M$ denote
the discriminant locus of $W$. We see that it typically has
numerous components which can cross along higher codimensional
loci in the moduli space. At a generic point $p$ on any given component of
the discriminant locus,
some collection of three-cycles have collapsed to zero period.
The mirror of this statement is that some collection of
{\it even}-dimensional cycles on $M$ have collapsed to zero quantum volume.
Notice that in general these are not just two-cycles. The precise
identification of which even-dimensional cycles have collapsed
can be determined from a detailed application of mirror symmetry
\ref\rGPM{B. Greene, R. Plesser and  D. Morrison, {\it Mirror Manifolds
in  Higher Dimension,} Commun.Math.Phys. {\bf 173} (1995) 559}
\ref\rMorrison{D. Morrison, {\it Making Enumerative Predictions by
Means of Mirror Symmetry}, in {\it Essays on Mirror Manifolds II},
(B. Greene, S.T. Yau, editors), International Press, 1996}.
Namely, in the neighborhood of maximal unipotent monodromy,
the even cycles in $H_{2j}(M,\BZ)$ are mirror to the quotient space
$W_{j+1}/W_{j}$ where the $W_{j}$ form the monodromy weight filtration
of $H_3(M,\BZ)$. Roughly, the three-cycles in $W_{j+1}/W_{j}$ have
periods with $log^j$ type monodromy about the maximal unipotent point.
More precisely, this statement is true up to lower order $log$ monodromy
transformations whose precise form requires more detailed study
of the mirror map than we shall undertake here. 

As we move along the component of the discriminant locus
on which $p$ lies, generically the homology class of the collapsed
three-cycles --- being  integral classes --- will remain constant
\foot{We thank M. Gross for discussions on this point.}. 
By construction, then, the homology class of the collapsed
even-cycles on $M$ will stay constant as well. The value
of $K$ along this component, though, will generally {\it change}.
Recall that 
$K$  depends upon the cycles with log-monodromy periods and hence
only if such a cycle is amongst those which have collapsed on this
component of the discriminant locus will the value of $K$, projected
onto that cycle, remain fixed at zero.

As an example, let's consider the case of the flop studied in
section 2. Using the picture just presented,
 the flopping two-cycle $S^2_{(j)}$
on $M$ is mirror to a log-monodromy period 
three-cycle $\gamma_j$ on $W$. Along
a component of the discriminant locus of $W$ on which this three-cycle
has zero period (such as the component containing 
point $\tilde{p}$ in Fig. 5), 
the value of $t_j$ will be identically zero. Appropriately
wrapped three-branes on $W$ and wrapped two-branes on $M$ will be massless.
This would give a general explanation of what we found earlier:
namely, flopped curves have
zero quantum volume (as measured by string instantons) at the flop-point,
regardless of the values of other K\"ahler moduli.

Along other components of the discriminant locus of $W$,
this picture  leads to the somewhat strange conclusion 
mentioned in the introduction
regarding the collapse of a cycle vs. the collapse of its subcycles.
Namely, if there are for instance components of the
discriminant locus of $W$
along which, say, the only vanishing three-cycles have  periods 
wtih $log^j , j > 1$ type monodromy
(up to lower order terms, with respect to  a maximal 
unipotent monodromy point),
then in the mirror description we will have higher dimensional
cycles collapsing to zero quantum volume while their lower dimensional
subcycles maintain nonzero quantum volume. As we move along this component
of the discriminant locus, the periods and identities of the
collapsed integral three-cycles, will be unchanged. However, in general
all other periods will be nonconstant and hence the quantum volumes of
all other cycles on the mirror $M$ will typically change. This includes
the quantum volumes of the subcyles of the collapsed cycles.
More generally, if $A$ is a subcycle of $B$ on $M$, then $B$ can collapse
to zero volume while $A$ does not so long as the mirror period
over the mirror three-cycle of $B$ on  $W$ vanishes, but that for $A$
does not. In any specific case, this is a question that can be  
answered by detailed
study of the discriminant locus of $W$.

The simplest explicit example of this phenomenon is given by
the quintic hypersurface in $\IC\IP^4$ and $W$, its mirror.
The discriminant locus in this case consists of a single point.
As explained by Candelas, et. al \rCan\  and utilized by
Strominger \rStrominger, at the point in $W$
\eqn\eW{z_1^5 + z_2^5 + z_3^5 + z_4^5 + z_5^5 - 5 \psi z_1z_2z_3z_4z_5 = 0}
with $\psi = 1$, one of the four three-cycles in $H_3(W,\BZ)$ collapses.
The expectation initially indicated in \rStrominger\ is that
the mirror to this
three-cycle in $M$ is a collapsing two-cycle. This, however, is not true
\foot{After completing this work, it was brought to our attention
that A. Strominger and J. Polchinski
\ref\rSP{ J. Polchinski and  A. Strominger, {\it
 New Vacua for Type II String Theory}, Phys.Lett. B388 (1996) 736.}
 have previously made the same
observation.}.
$M$ has a single K\"ahler modulus with component $t_1 = B_1 + i J_1$
with respect to the integral generator of $H_2(M,\BZ)$. At the mirror
to the $\psi = 1$ point, the value of $t_1$ is nonzero: $t_1 \approx 1.2056$
(the normalization factor $N$ is also finite at this point). Thus,
the two-cycle has not collapsed: $D$-2-branes wrapping the two cycle
will not become massless. Rather, using the monodromies calculated
in \rCan, it is not hard to see that the vanishing three-cycle at
$\psi = 1$, is in the homology class of the three-cycle which at
infinity has period with $log^3$-type monodromy. Thus, up to lower order
terms, it is the {\it six}-cycle on $M$ which collapses at the mirror
of the $\psi = 1$ point. Notice that in this example we only have one
K\"ahler modulus at our disposal and by tuning it appropriately the
quantum volume of the whole Calabi-Yau vanishes even though the quantum
volume of the homology two-cycle does not. This accounts for
us getting the desired result of a single new massless particle,
even though the entire space is collapsing.

It would be of interest to further explore the properties of
quantum volume we have defined. Situations in which four-cycles
collapse in various ways \rSeibergdelPezzo \rVafadelPezzo\
 are prime examples in which the behavior
of sub-two-cycles is important to the resulting physics.
Preliminary study has shown that at least in some examples,
Calabi-Yau transitions involving collapsing del Pezzo surfaces
involve cycles with nonzero quantum volumes. This should
allow for the verification of the conjecture made at the
end of \rVafadelPezzo.  We will report
on this elsewhere.

\vskip.2in
{\centerline {\bf Acknowledgments}}

We are happy to acknowledge a number of useful discussions with
Paul Aspinwall, Mark Gross, Calin Lazaroiu, David Morrison and Cumrun Vafa.
BRG is supported in part by the National Science Foundation,
a National Young Investigator Award and the Alfred P. Sloan Foundation.
YK is supported in part by the National Science Foundation.

\listrefs

\end